\documentclass[conference]{IEEEtran}
\makeatletter
\def\ps@headings{%
\def\@oddhead{\mbox{}\scriptsize\rightmark \hfil \thepage}%
\def\@evenhead{\scriptsize\thepage \hfil \leftmark\mbox{}}%
\def\@oddfoot{}%
\def\@evenfoot{}}
\makeatother
\usepackage{graphicx,cite}
\usepackage{epsfig}
\usepackage[cmex10]{amsmath}

\usepackage{amssymb}
\usepackage{array}
\usepackage{fixltx2e}
\usepackage{amsfonts}
\usepackage{lineno}

\begin{document}
\title{Optimal Resource Allocation and Relay Selection in Bandwidth Exchange 
Based Cooperative Forwarding}

\author{\IEEEauthorblockN{Muhammad Nazmul Islam\IEEEauthorrefmark{0}, 
Narayan Mandayam\IEEEauthorrefmark{0} 
 and Sastry Kompella\IEEEauthorrefmark{1}  \\
 }
 \IEEEauthorblockA{\IEEEauthorrefmark{0}
%Dept.~of Elec.~and Comp.~Eng.,
WINLAB, Rutgers University, Email: mnislam@winlab.rutgers.edu, narayan@winlab.rutgers.edu \\
\IEEEauthorrefmark{1} Information Technology Division,
Naval Research Laboratory, Email: sk@ieee.org}}

\maketitle

\begin{abstract}

In this paper, we investigate joint optimal relay selection and resource allocation
under bandwidth exchange (BE) enabled
incentivized cooperative forwarding in wireless networks. We consider an autonomous network where $N$ nodes
transmit data in the uplink to an access point (AP) / base station (BS). We consider the
scenario where each node gets an initial amount (equal, optimal based on direct path or arbitrary) of bandwidth,
and uses this bandwidth as a flexible incentive for two hop relaying. 
We focus on $\alpha$-fair network utility maximization (NUM)
and outage reduction in this environment. 
Our contribution is two-fold. First, we propose an incentivized forwarding based
resource allocation algorithm which maximizes the global utility while preserving the initial utility of
each cooperative node. Second, defining the link weight 
of each relay pair as the utility gain due to cooperation (over noncooperation), 
we show that the optimal relay selection in $\alpha$-fair NUM
reduces to the maximum weighted matching (MWM) problem in a non-bipartite graph.
% Third, we show that a similar
%reduction to maximum matching (MM) in a bipartite graph is possible for the case of outage probability reduction.
Numerical results show that the proposed algorithms provide 20-25\% gain in spectral efficiency
and 90-98\% reduction in outage probability.
%significant improvements in terms of outage probability, 
%sumrate, maxmin rate and total power performance of the network.

\end{abstract}
%\begin{IEEEkeywords}
%Resource Allocation, Incentivized Forwarding, Relay Selection, Maximum Weighted Matching.
%\end{IEEEkeywords}

\section{Introduction}

The benefits of cooperative communications~\cite{Tse} have led to significant research 
in relaying and forwarding~\cite{ElGamal}. 
%The benefits of cooperative communications is well known. 
However, forwarding always incurs some costs, e.g., power and/or delay.
There have been few works that focus on the explicit cost of forwarding.
%A forwarder node's data rate may drop if it has to relay the sender node's data.
Existing cooperative communications literature include several 
incentive based mechanisms to encourage the forwarder nodes for cooperation.  
%These include pricing based mechanisms~\cite{Ileri}, reputation based mechanisms and
%mechanisms based on forwarding games~\cite{Buttyan}.
These techniques include pricing~\cite{Ileri}, reputation~\cite{Buchegger} 
and credit~\cite{Buttyan} based cooperative forwarding.
However, these mechanisms require a stable economy or a shared understanding of what
things are worth and become unrealizable in dynamic wireless networks.

In light of this, the authors of~\cite{Zhang} developed 
a bandwidth exchange (BE) enabled incentive mechanism where
nodes offer a portion of their allocated bandwidths to other nodes as immediate incentives for relaying.
They used a Nash bargaining solution based resource allocation and a heuristic relay
selection policy in their work. In this work, we focus on the distributed joint optimal relay selection
and resource allocation in the $\alpha$-fair NUM and outage probability reduction 
of a BE enabled network. 

We consider an $N$ node autonomous network where each node receives an initial amount 
(equal, optimal based on direct path transmission or arbitrary) of bandwidth and 
connects directly to the access point (AP) / base station (BS). We consider a frequency division multiple access system where all nodes transmit at the same time with different bandwidth slots.
In this context, we focus on a two-hop incentivized cooperative forwarding scheme where a sender
node provides bandwidth as an incentive to a forwarder node for relaying its data
to the AP/BS. 
%We assume one sender for one forwarder and pairwise total bandwidth constraint, i.e.,
%the cooperation between one pair of nodes does not change the bandwidth allocation in other nodes. 
We first prove the concavity of the resource allocation problem and
then show that the optimal relay selection problem in $\alpha$-fair NUM
reduces to the classical non-bipartite maximum weighted matching (MWM) algorithm~\cite{Edmonds}.
%At first, we prove the concavity of the resource allocation problem formulation.
%Thereafter, defining the link weights as the pairwise difference between cooperation and non-cooperation
%utility, we show that the optimal relay selection problem reduces to the classical
%nonbipartite maximum weighted matching (MWM) problem~\cite{Edmonds,Preis}. 
Using the distributed local greedy MWM~\cite{Hoepman},
we propose a simple distributed BE enabled incentivized forwarding protocol. 
We also show that the outage probability reduction problem reduces to the 
bipartite maximum matching algorithm in this context.
Numerical simulations show that the proposed algorithm provides 20-25\% spectrum
efficiency gain and 90-98\% outage probability reduction in a $20$ node network.

%simulations also illustrate that no coopeartive node's rate drops 
%below its initial rate in the proposed algorithm.
%Numerical simulations of 
%The full version of the paper, if accepted, will show 
%the proposed algorithm's contribution in 
%transmission power reduction. It will also illustrate
%the application of time exchange (TE) enabled incentivized forwarding
%in a time division multiple access (TDMA) network.

\subsection{Related Work \& Our Contributions}

Our contributions in this paper can be summarized as follows.
First, we consider incentivized relaying in a network
where each node has been allocated an initial amount of bandwidth. Previously,
the authors of~\cite{Zhang2} and~\cite{Baochun:a} considered incentivized
forwarding  in a cognitive radio network where 
only the primary users initially receive resources and later transfer 
some of their resources to the secondary users as incentives for relaying.
%The authors of~\cite{Zhang2,Baochun:a} also assumed that each primary user knows the 
%channel information of all secondary users. On the other hand, our distributed incentivized
%protocol only requires the local knowledge of neighbouring node's channel information.
In contrast, our work focuses on distributed incentivized two-hop relaying in an
autonomous network where a centralized algorithm
%, based on all two-hop path information,
might be infeasible due to the associated long estimation delay and high complexity.

Second, our proposed decode \& forward (DF) BE 
enabled resource allocation maximizes the summation of the utilities  
while preserving the initial utilities of the individual nodes.
Previously, the authors of~\cite{Nazmul} considered BE from a simpler two hop relaying
perspective. The authors of~\cite{Foschini} proposed a similar half duplex DF relaying approach. However, they considered a commercial relay network
where the relay did not have its own data~\cite{Foschini}.
To the best of our knowledge, our proposed BE based resource allocation algorithm has not
been investigated before.

%Third, we define the link weight of each cooperative pair as the difference 
%between cooperation and non-cooperation utility. This definition of link weight
%along with the one forwarder for one sender and pairwise resource constaint assumptions 
%makes the optimal relay selection problem converge to the nonbipartite MWM problem.
%Note that, 
Third, our definition of link weight in the use of MWM 
is different from that of existing literature.
Inspired by the seminal work on maximum weighted scheduling~\cite{Ephremides}, most of the works 
on MWM based scheduling
%have focused on stability properties of a throughput maximized network and 
have defined link weights as the differential backlog size of that particular link~\cite{Neely,Saswati}.
%The authors of~\cite{Neely} derived the MWM based joint scheduling, routing
%and power allocation policy that achieves the maximum network layer capcity.
%In these works, each node contains data that needs 
%to be sent to some other nodes in the network. 
%The differential backlog of a link is the difference
%between the number of packets waiting at the source node and that at the 
%destination node~\cite{Saswati}. 
Based on this definition and using the network layer 
capacity perspective, the MWM algorithm of these works finds
the set of links that will be activated at each slot~\cite{Chiang:b,Shroff}. However,  
%all nodes talk to the AP together at the same time with different bandwidth slots. 
we adopt an information theoretic capacity perspective in this work. We
define the link weights of the MWM graph as the utility gain that 
a DF relaying enabled cooperative pair offers to the system.
Therefore, the `matched' nodes of the MWM algorithm communicate to the AP using a DF cooperation 
strategy whereas, the `unmatched' nodes transmit to the AP 
%in the direct path 
without cooperating with any other node. 
In this regards, our relay selection approach is closer to the work of~\cite{Mak}
where the authors defined link weights of each cooperative pair as the energy savings of cooperation
over noncooperation. However,~\cite{Mak} considered energy minimization from a bit error rate (BER)
perspective whereas, we focus on $\alpha$-fair NUM from capacity perspective.

This paper is organized as follows. Section~\ref{sec:Model} and~\ref{sec:Objective} show the proposed incentivized system model
and the design objective respectively.
Section~\ref{sec:Incentive} describes the problem formulations and solutions of incentivized forwarding based resource
allocation and relay selection.
After providing the simulations results in Section~\ref{sec:Simulation}, we conclude 
the work in Section~\ref{sec:Conclusion}.

\begin{figure}[t]
%\vspace{0.8cm}
%\begin{minipage}[b]{0.5\linewidth}
\centering
\includegraphics[scale=0.35]{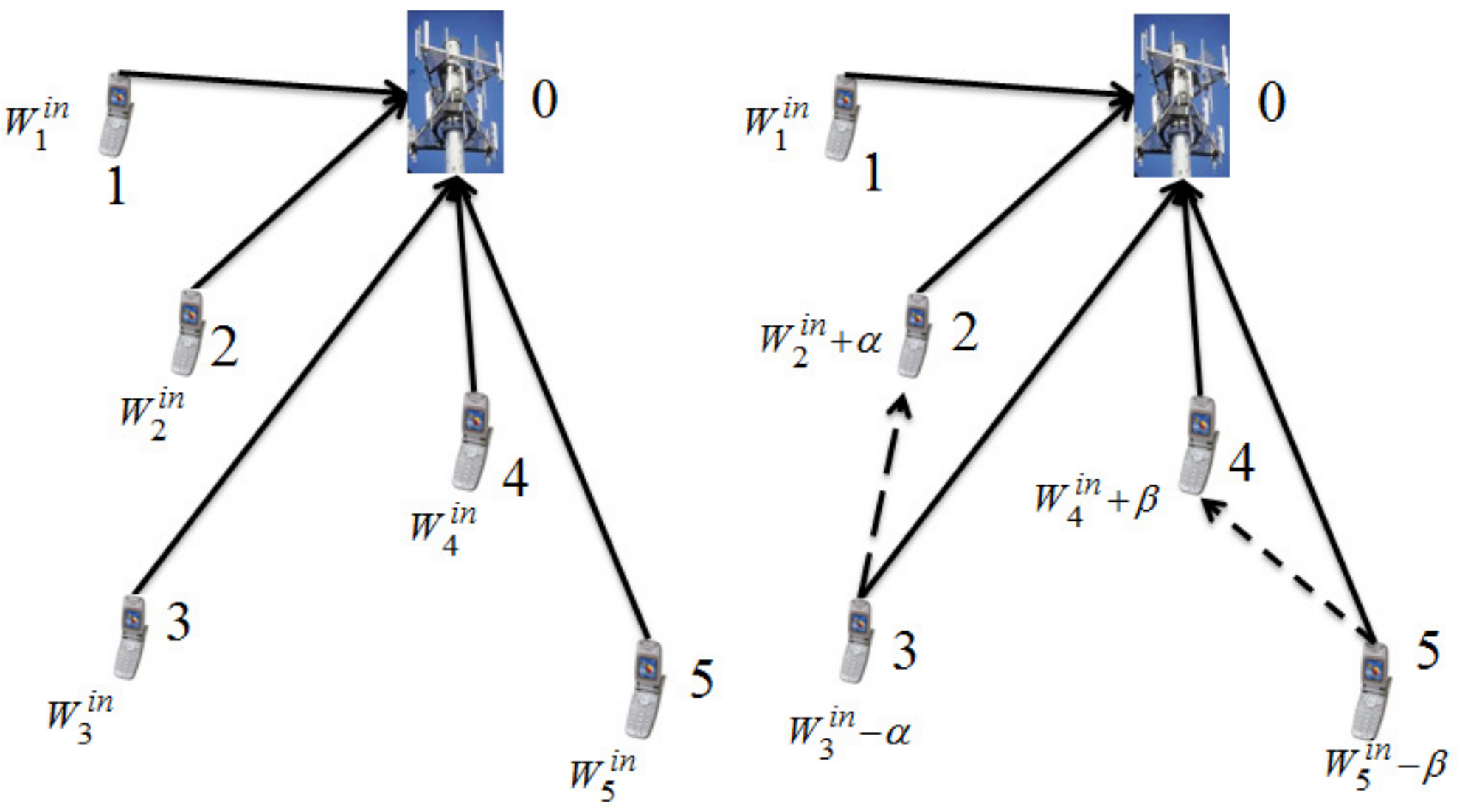}
\caption{Direct Transmission and BE enabled Incentivized Forwarding}
\centering
\label{fig:BEModel}
\end{figure}
%\end{minipage}
%\hspace{1cm}
%\begin{minipage}[b]{0.5\linewidth}

\section{System Model}   \label{sec:Model}

We consider the uplink of an $N$ node single cell FDMA network.
Let $\mathcal{V} = \{1, \, 2, \, \cdots, \, N \}$ denote the set of $N$ nodes
that transmit data to the BS (node $0$).
Each node uses the total time slot. Node 
$i \, \in \, \mathcal{V}$ is initially allotted a bandwidth of $W_i^{in}$. 
Let $\rho_{ij}$ and $R_{ij}$ denote the link
gain and achievable rate of the $ij$ path respectively. For a given node $i \in [1, N]$
with $W_i^{in}$ bandwidth and direct link gain $\rho_{i0}$, the achievable throughput is:
\begin{equation}
R_i^{in} = R_{i0} = W_i^{in} \log_2 \bigl(1 + \frac{\rho_{ij} P_{i,max}}{W_i^{in}} \bigr) 
\label{eq:DirectRate}
\end{equation}
Here, $P_{i,max}$ denotes the maximum transmission power of node $i$ and $R_i^{in}$
is expressed in bit per second (bps). 

In BE, nodes perform two hop half duplex DF cooperative relaying.
The forwarder node hears the sender nodes' data and relays some of that data
along with transmitting its own data to the AP. Since each node is a power constrained
device, the forwarder nodes' data rate may drop if it continues to transmit with the
same bandwidth. Therefore, the sender node delegates some of its bandwidth
to the forwarder node as incentives for relaying. 
We consider one forwarder for one sender and vice versa to reduce the relay searching complexity.
Let $ \mathcal{SF} = \{\mathcal{SF}_1,\cdots,\mathcal{SF}_K \} =
\{(s_1,f_1),(s_2,f_2),\cdots,(s_K,f_K)\}$ denote the sender-forwarder
pair set, i.e., $f_i$ relays $s_i$'s data along with transmitting $f_i$'s own data. 
Let $\mathcal{D} = {d_1, d_2, \cdots, d_L}$ denote the direct set, i.e., the 
set of remaining nodes that transmit data without cooperation. 
%Let $ |\mathcal{SF}| = 2*K$. $|\mathcal{D}| = L$.
Note that, $K$ and $L$ are variables and further,  $2* K + L = N$.
We assume pairwise bandwidth constraint in this work.

The left and right sides of Fig.~\ref{fig:BEModel} show the considered direct transmission 
model and the proposed BE model. In BE, node $2$ relays data for node $3$. 
Node $3$ delegates $\alpha$ amount of bandwidth to node $2$ as incentive for relaying. 
Let $W_i^{be}$ represent the bandwidth of node $i$ in the BE scenario. 
Now, $W_2^{be} = W_2^{in} + \alpha$ and $W_3^{be} = W_3^{in} - \alpha$. 
Node $4$ and $5$ operate in the same manner.

Nodes don't do power allocation among different streams in our framework.
Since capacity is a non-decreasing function of transmission power, 
each node utilizes the maximum transmission power in its allotted bandwidth slot.
Without loss of generality, we assume $P_{i,max} = P \, \forall \, i \, \in \, \mathcal{V}$ 
in the subsequent analysis.

\begin{figure}[t]
%\vspace{0.8cm}
\centering
\includegraphics[scale=0.37]{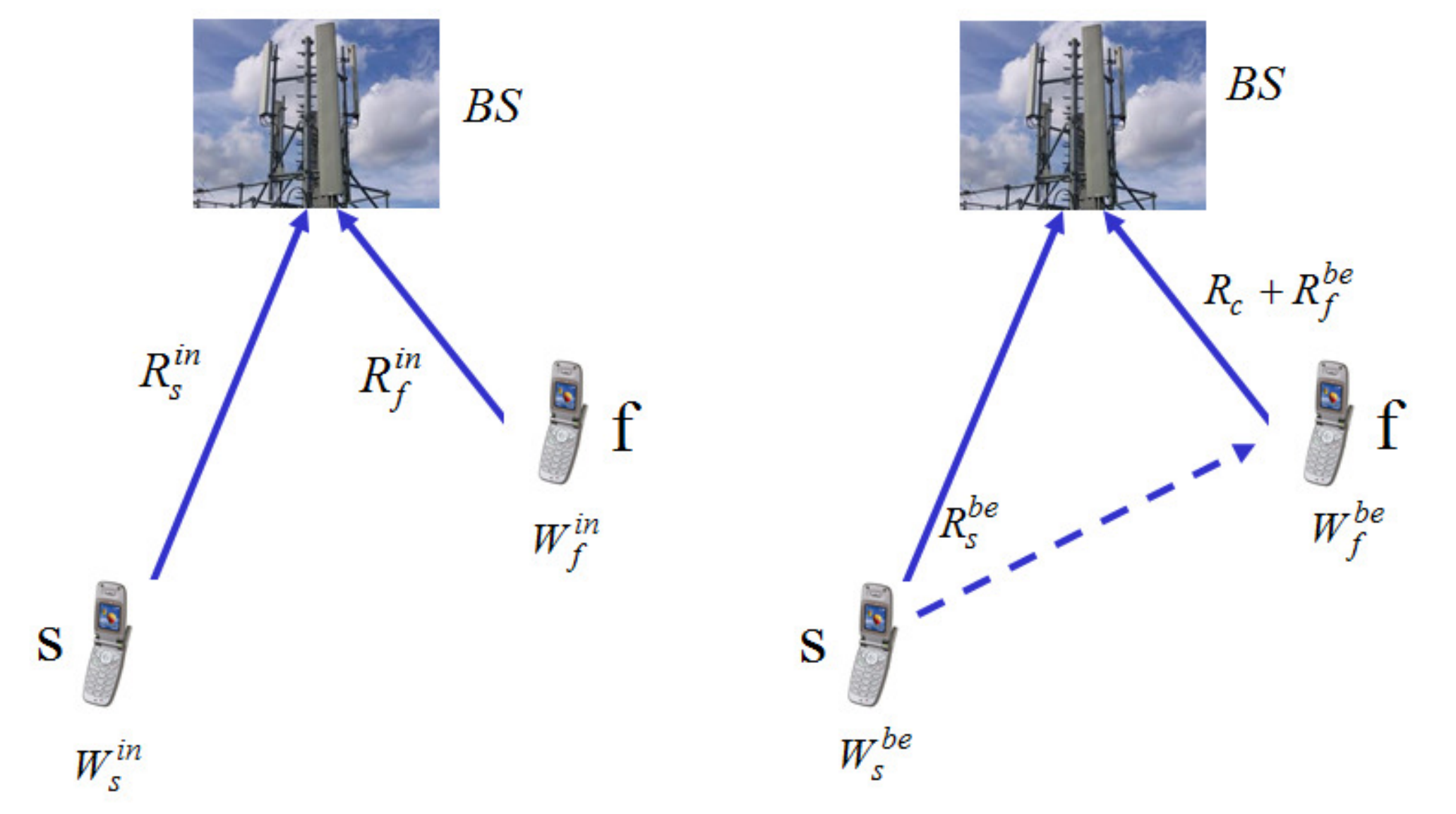}
\caption{BE Enabled Forwarding in a $3$ Node Network}
\centering
\label{fig:BETwoNode}
%\end{minipage}
\end{figure}

\subsubsection{Rate analysis in the BE scenario: }

\begin{table}
\caption{Summary of used notations} \label{Tab:one}
\begin{center}
\begin{tabular}{|l|l|} \hline
Notation & Meaning \\ \hline
$N$ & Total number of users \\ \hline
%$W$ & Transmission bandwidth \\ \hline
%$B$ & Total availble bandwidth \\ \hline
%$T$ & Transmission time \\ \hline
$P$ & Maximum transmission power \\ \hline
%$S$ & Instantaneous transmission power \\ \hline
$\rho_{ij}$ & Gain of the $ij$ link \\ \hline
%$C$ & Channel capacity  \\ \hline
$W_i^{in}$ & Initial bandwidth of node $i$ \\ \hline
$W_i^{be}$ & Node $i$'s bandwidth in BE \\ \hline
$R_i^{in}$ & Initial rate of node $i$ \\ \hline
$R_i^{be}$ & Node $i$'s rate in BE \\ \hline
$R_{ij}$ & Achievable rate in the $ij$ link \\ \hline
$\mathcal{V}$ & Set of all nodes \\ \hline
%$\mathcal{E}$ & Set of graph edges \\ \hline
%$\mathcal{G}$ & Undirected graph, $\mathcal{G} = (\mathcal{V}, \mathcal{E})$ \\ \hline
$\mathcal{D}$ & Set of nodes that transmit without cooperation  \\ \hline
% treceive data through direct path
%$\mathcal{S}$ & Set of sender nodes \\ \hline
%$\mathcal{F}$ & Set of forwarder nodes \\ \hline
$\mathcal{SF}$ & Set of sender-forwarder pairs \\ \hline
$SF_i$ & $i^{th}$ sender-forwarder pair $(s_i,f_i)$ \\ \hline
\end{tabular}
\end{center}
\end{table}

Let $R_i^{be}$ represent the achievable rate of node $i$ in the 
BE scenario. The right side of Fig.~\ref{fig:BETwoNode} 
shows the interaction between a sender node $s$
and a forwarder node $f$. 
%Here, $R_r$ and $R_d$ denote the 
%achievable rates in the sender-forwarder path and sender-BS path respectively.
%In other words, $R_r = R_{sf}$, $R_d = R_{s0}$. 
If node $s$ transmits to node $f$ and the BS \emph{separately} using $W_s^{be}$ bandwidth, 
the achievable throughput in the respective paths are:
\begin{eqnarray}
R_{sf} & = & W_s^{be} \log_2 \bigl(1 + \frac{\rho_{sf} P}{W_s^{be}} \bigr)  \label{eq:RateSF}\\
R_{s0} & = & W_s^{be} \log_2 \bigl(1 + \frac{\rho_{s0} P}{W_s^{be}} \bigr)   \label{eq:RateS0}
\end{eqnarray}
If $f$ transmits to the BS using $W_f^{be}$ bandwidth,
\begin{equation}
R_{f0} = W_f^{be} \log_2 \bigl(1 + \frac{\rho_{f0} P}{W_f^{be}} \bigr)   \label{eq:RateF0}
\end{equation} 
Assuming $\rho_{sf} \geq \rho_{s0}$, it is easily seen that $R_{sf} \geq R_{s0}$. 
Due to the nature of the wireless environment, when sender $s$ transmits, both
$f$ and BS hear it. If node $s$ transmits at rate $R_{sf}$,
node $f$ can decode it properly. 
The BS also receives the same signal but can't decode it properly since 
$R_{sf} \geq R_{s0}$. However, node $f$ can forward $R_c$ bits to the BS 
to resolve the BS's uncertainty about node $s$'s data. Node $f$ 
also transmits its own data, $R_f^{be}$, to the BS. 
The information theoretic generalization of the 
maximum-flow-minimum-cut-theorem~\cite{Cover2} provides the following 
relationship between these achievable rates,
\begin{eqnarray}
R_s^{be} & \leq & \min (R_{sf}, R_{s0} + R_c)  \label{eq:RateSender}   \\
R_c + R_f^{be} & \leq & R_{f0}    \label{eq:RateForwarder}
\end{eqnarray}
The codebook design procedure to achieve these rates is summarized in Appendix~\ref{sec:Codebook}.
A detailed description can be found in~\cite{Cover2,Cover}. 

Using this rate analysis, we focus on the distributed joint optimal
bandwidth allocation and relay selection in maximizing the summation of 
the $\alpha$-fair utilities of the network.

\emph{$\alpha$-fair utility: }

The $\alpha$-fair utility is defined for any $ \alpha \, \in \, [0, \infty)$, as~\cite{Mo},
\begin{equation}
  U^{\alpha}(R)=\begin{cases}
    \frac{R^{1-\alpha}}{1-\alpha}, & \text{if \, $\alpha \neq 1$}\\
    log(R), & \text{if \, $\alpha = 1$}
  \end{cases}   \label{eq:AlphaFair}
\end{equation}
where $R$ represents the rate of the user.
Summation of $\alpha$-fair utility functions takes the form of different well known
utility functions, %~\cite{Chiang:a},
e.g., sum rate maximization ($\alpha = 0$), proportional fairness ($\alpha = 1$) and minimum rate
maximization ($\alpha = \infty$).

\section{Design Objective}   \label{sec:Objective}

We begin with a focus on $\alpha$-fair NUM of the overall network through optimal bandwidth 
and rate allocation for all possible sender-forwarder pairing sets.

%Our objective is to perform the joint optimal relay selection and bandwidth and rate allocation
%in the $\alpha$-fair NUM
% and outage probability reduction 
%of the BE enabled relay network. We, at first,
%provide the $\alpha$-fair NUM objective:

\emph{\newline \underline{Problem I}}
\begin{subequations}
\begin{equation}
%\underset{\mathcal{D}, \mathcal{SF}, R_s^{be}, R_f^{be}, W_s^{be}, W_f^{be}}
\mathbf{\max.}  \, \sum_{d \in \mathcal{D}} U^{\alpha} (R_d^{be}) +  \sum_{(s, \, f) \in \mathcal{SF}} 
\bigl(U^{\alpha} (R_f^{be})  + U^{\alpha} (R_s^{be}) \bigr)
 \label{eq:MainObjective}
\end{equation}
\begin{equation}
\mathbf{s.t.} \, \, (R_f^{be},R_s^{be}) \, \in \, conv(W_f^{be},W_s^{be}) 
\, \, \forall \, (s,f) \, \in \, \mathcal{SF}   \label{eq:RateConstraint1}  
\end{equation}
\begin{equation}
R_f^{be} \geq R_f^{in} \, , \, R_s^{be} \geq R_s^{in} 
\, \, \forall \, (s,f) \, \in \, \mathcal{SF}   \label{eq:RateConstraint2} 
\end{equation}
\begin{equation}
W_f^{be} + W_s^{be} \, \leq \, W_f^{in} + W_s^{in} \,  \, 
\forall \, (s,f) \, \in \, \mathcal{SF}  \label{eq:BandwidthConstraint1}
\end{equation}
\begin{equation}
\, \, W_f^{be}, \, W_s^{be} \, \geq 0 \, \, \forall \, (s,f) \, \in \, \mathcal{SF}  \label{eq:BandwidthConstraint2}
\end{equation}
\begin{equation}
\mathcal{D} \subseteq \mathcal{V} \, , \, \mathcal{SF} \in \mathcal{V} \times \mathcal{V} \, \, , \, \,
\mathcal{SF}_i \cap \mathcal{SF}_j = \emptyset  \, \, \forall i \neq j  \label{eq:RelaySelection1}
\end{equation}
\begin{equation}
\mathcal{SF}_i \cap \mathcal{D} = \emptyset \, \forall \, i \, \in [1,K]  
\label{eq:RelaySelection2}
\end{equation}
\begin{equation}
\mathcal{SF}_1 \cup \mathcal{SF}_2 \cdots \cup \mathcal{SF}_K \cup \mathcal{D} = \mathcal{V} 
\label{eq:RelaySelection3}
\end{equation}
\begin{equation}
Variables \, \, \mathcal{D}, \mathcal{SF}, R_s^{be}, R_f^{be}, W_s^{be}, W_f^{be}
\nonumber
\end{equation}
%\begin{equation}
%\mathcal{SF}_i \cap \mathcal{SF}_j = \emptyset  \, \, \forall i \neq j \, , \, \mathcal{S} \cap \mathcal{F}
%= \mathcal{S} \cap \mathcal{D} = \mathcal{F} \cap \mathcal{D} = \emptyset \, , \, \mathcal{S} \cup \mathcal{F} \cup \mathcal{D} = \mathcal{V} 
%\label{eq:RelaySelection2}
%\end{equation}
\end{subequations}
Here, $R_d^{be} = R_d^{in} \, \forall \, d \, \in \, D$. 
Therefore, the rates of the direct node set are not
optimization variables. Equation~\eqref{eq:RateConstraint1} 
denotes that the rate of the sender and forwarder lie in the 
convex hull of the allocated bandwidth. The details of this convex hull has already been
explained in the system model. It will also be mentioned in the next section.
Eq.~\eqref{eq:RateConstraint2} represents that the sender and the forwarders'
rate cannot drop below their initial rates. 
Equation~\eqref{eq:BandwidthConstraint1} shows that 
the total bandwidth used by the cooperative pair is constrained
by the summation of the initial bandwidths allocated to the individual nodes. Equation \eqref{eq:RelaySelection1}-~\eqref{eq:RelaySelection3} represent
the relay selection constraints. Equation \eqref{eq:RelaySelection1} 
shows that the direct and sender-forwarder sets are all subsets of the overall set. 
%Here, $SF_i$ represents the $i^{th}$ sender-forwarder pair, i.e., $SF_i = (s_i,f_i)$.
Eq.~\eqref{eq:RelaySelection2} denotes that 
the sender-forwarder pairs and direct set cannot have any common nodes. 
Eq.~\eqref{eq:RelaySelection3} represents that the
union of the pairs and the direct set form the overall set $\mathcal{V}$.
  
The solution of the above optimization problem depends 
on the selected sender-forwarder and direct node set
and the corresponding bandwidth and rate allocations. Hence, it
involves an exponential number of variables and constraints. In the rest of the paper, 
we focus on solving this problem.

\section{Optimization Problem Solution}   \label{sec:Incentive}

\subsection{Modified Optimization Problem}

Let $U^{\alpha} (R_{tot}) = \sum_{i \in \mathcal{V}} U^{\alpha} (R_i^{in})$ denote the 
summation of the initial utilities of the nodes. 
For a fixed $\mathcal{SF}$, $U^{\alpha} (R_{tot})$ can be
expressed in the following form:
\begin{eqnarray}
& & U^{\alpha} (R_{tot})  \nonumber \\
& = & \sum_{i \in \mathcal{V}} U^{\alpha} (R_i^{in})  \nonumber \\
& = & \sum_{d \in \mathcal{D}} U^{\alpha} (R_d^{in}) +  \sum_{(s, \, f) \in \mathcal{SF}} 
\bigl(U^{\alpha} (R_f^{in})  + U^{\alpha} (R_s^{in}) \bigr)  \label{eq:Modified1}   \\
& = & \sum_{d \in \mathcal{D}} U^{\alpha} (R_d^{be}) +  \sum_{(s, \, f) \in \mathcal{SF}} 
\bigl(U^{\alpha} (R_f^{in})  + U^{\alpha} (R_s^{in}) \bigr)  \label{eq:Modified2} 
\end{eqnarray}
Equation~\eqref{eq:Modified1} follows from eq.~\eqref{eq:RelaySelection3}.
Equation~\eqref{eq:Modified2} uses the fact that $R_d^{be} = R_d^{in} \, \forall \, d \in D$.
Subtracting $U^{\alpha} (R_{tot})$ from the objective function of I, we find
the following optimization problem:

\emph{\newline \underline{Problem II}}
\begin{subequations}
\begin{equation}
 \sum_{(s, \, f) \in \mathcal{SF}} 
\bigl( U^{\alpha} (R_f^{be})  + U^{\alpha} (R_s^{be}) - 
U^{\alpha} (R_f^{in})  + U^{\alpha} (R_s^{in}) \bigr)
 \label{eq:ModifiedObjective}
\end{equation}
\begin{equation}
\mathbf{s.t.} \, \, (R_f^{be},R_s^{be}) \, \in \, conv(W_f^{be},W_s^{be}) 
\, \, \forall \, (s,f) \, \in \, \mathcal{SF}    
\end{equation}
\begin{equation}
R_f^{be} \geq R_f^{in} \, , \, R_s^{be} \geq R_s^{in} 
\, \, \forall \, (s,f) \, \in \, \mathcal{SF}    
\end{equation}
\begin{equation}
W_f^{be} + W_s^{be} \, \leq \, W_f^{in} + W_s^{in} \,  \, 
\forall \, (s,f) \, \in \, \mathcal{SF}  
\end{equation}
\begin{equation}
\, \, W_f^{be}, \, W_s^{be} \, \geq 0 \, \, \forall \, (s,f) \, \in \, \mathcal{SF} 
\end{equation}
\begin{equation}
\mathcal{D} \subseteq \mathcal{V} \, , \, \mathcal{SF} \in \mathcal{V} \times \mathcal{V} \, \, , \, \,
\mathcal{SF}_i \cap \mathcal{SF}_j = \emptyset  \, \, \forall i \neq j  
\end{equation}
\begin{equation}
\mathcal{SF}_i \cap \mathcal{D} = \emptyset \, \forall \, i \, \in [1,K]  
\end{equation}
\begin{equation}
\mathcal{SF}_1 \cup \mathcal{SF}_2 \cdots \cup \mathcal{SF}_K \cup \mathcal{D} = \mathcal{V} 
\end{equation}
\begin{equation}
Variables \, \, \mathcal{D}, \mathcal{SF}, R_s^{be}, R_f^{be}, W_s^{be}, W_f^{be}
\nonumber
\end{equation}
\end{subequations}
The inclusion of constant terms in the objective function does not change 
the optimal variables of an optimization problem~\cite{Boyd}. 
%Therefore, the set of optimal variables of both problem I and problem II are same.
As a result, the same set of sender-forwarder pairs maximize both problem
I and II. We will focus on solving problem II in the subsequent analysis.
The optimal variables of problem II will directly lead to the optimal solution
of problem I.

Problem II depends on both relay selection and resource
allocation. The very nature of the design objective allows us to
split the optimization formulation into the following two parts:
\begin{itemize}
\item For any fixed set of sender-forwarder pairs, perform DF based optimal rate and bandwidth allocation.
\item Choose the relay set that maximizes the summation of the $\alpha$-fair utility of the nodes.
\end{itemize}
We, at first, focus on resource allocation in a fixed sender-forwarder pair set $\mathcal{SF}$
and direct node set $\mathcal{D}$. Later, we will show the optimal sender-forwarder pair 
selection policy.

\subsection{Optimal bandwidth and rate allocation for a fixed sender-forwarder set}

The optimal resource allocation problem for fixed 
$\mathcal{SF}$ and $\mathcal{D}$ takes the following form:

%\emph{\newline \underline{Problem II}}
%%
%\begin{subequations}
%\begin{equation}
%\underset{R_s^{be}, R_f^{be}, W_s^{be}, W_f^{be}}{\mathbf{\max.}} \, \, \sum_{d \in \mathcal{D}} U^{\alpha} (R_d^{be}) +  \sum_{(s, \, f) \in \mathcal{SF}} 
%\bigl(U^{\alpha} (R_f^{be})  + U^{\alpha} (R_s^{be}) \bigr)
% \label{eq:SubObjective1}
%\end{equation}
%\begin{equation}
%\mathbf{s.t.} \, \, (R_f^{be},R_s^{be}) \, \in \, conv(W_f^{be},W_s^{be}) 
%\, \, \forall \, (s,f) \, \in \, \mathcal{SF}   \label{eq:SubRateConstraint1}  
%\end{equation}
%\begin{equation}
%R_f^{be} \geq R_f^{in} \, , \, R_s^{be} \geq R_s^{in} 
% \, \, \forall \, (s,f) \, \in \, \mathcal{SF} 
% \label{eq:SubRateConstraint2} 
%\end{equation}
%\begin{equation}
%W_f^{be} + W_s^{be} \, \leq \, W_f^{in} + W_s^{in} \,  \, , \, \, W_f^{be}, \, W_s^{be} \, \geq 0 \, \, \forall \, (s,f) \, \in \, \mathcal{SF}  \label{eq:SubBandwidthConstraint1}
%\end{equation}
%\end{subequations}
%%
%Let $U^{\alpha} (R_{tot}) = \sum_{i=1}^N U^{\alpha} (R_i^{in})$ denote the 
%summation of the initial utilities of the nodes. 
%%$U^{\alpha} (R_{tot})$ is a constant with respect to the variables of problem II. tH
%% Summing $U^{\alpha} (R_i^{in})$ for all nodes and 
%Subtracting $U^{\alpha} (R_{tot})$ from the objective function of II, we find: 
\emph{\newline \underline{Problem III}} 
\begin{subequations}
\begin{equation}
 \sum_{(s, \, f) \in \mathcal{SF}} 
\bigl( U^{\alpha} (R_f^{be})  + U^{\alpha} (R_s^{be}) - U^{\alpha} (R_f^{in})  - U^{\alpha} (R_s^{in}) \bigr)  
 \label{eq:DiffObjective1}
\end{equation}
\begin{equation}
\mathbf{s.t.}  \, (R_f^{be},R_s^{be}) \, \in \, conv(W_f^{be},W_s^{be}) 
 \, \, \forall \, (s,f) \, \in \, \mathcal{SF}   \label{eq:DiffRateConstraint}  
\end{equation}
\begin{equation}
 R_f^{be} \geq R_f^{in} \, \, , \, \, R_s^{be} \geq R_s^{in} 
\, \, \, \forall \, (s,f) \, \in \, \mathcal{SF} 
\label{eq:DiffRateConstraint2}
\end{equation}
\begin{equation}
W_f^{be} + W_s^{be} \, \leq \, W_f^{in} + W_s^{in} \,  \, , \, \, W_f^{be}, \, W_s^{be} \, \geq 0 \, \, \forall \, (s,f) \, \in \, \mathcal{SF}  \label{eq:DiffBandwidthConstraint1}
\end{equation}
\begin{equation}
Variables \, \, R_s^{be},\, R_f^{be},\, W_s^{be},\, W_f^{be}
\end{equation}
\end{subequations}
%
%The inclusion of constant terms in the objective function does not change 
%the optimal variables of an optimization problem~\cite{Boyd}. 
%Therefore, the set of optimal variables of both problem II and problem III are same.
%
%The nodes in $\mathcal{D}$, i.e., the direct set, transmit at same data rate in both BE and initial mode. Therefore, these terms get cancelled
%after subtraction and don't appear in~\eqref{eq:DiffObjective1}. 
% The initial data rates
%, e.g. $R_s^{in}$ and $R_f^{in}$ are constants with respect to the BE based resource allocation.
% Hence, these are not optimization variables. 
%the presence of in~\eqref{eq:DiffObjective1} does not change the optimal variables.
Now, due to the pairwise bandwidth constraint of~\eqref{eq:DiffBandwidthConstraint1}, the bandwidth allocation in one cooperative pair does not affect other nodes. Therefore, problem III is just the summation of $K$ independent three node (sender, forwarder
and BS) resource allocation problems and can be decomposed into the subproblems.
Hence, we now focus on an arbitrary sender-forwarder pair $(s,f)$ and describe
the resource allocation problem formulation in this pair.
% by providing details on the convex hull of the achievable rates.

%Fig.~\ref{fig:BETwoNode} shows the interaction of an arbitrary sender-forwarder pair.
%Initially, nodes $s$ and $f$ receive $W_s^{in}$ and $W_f^{in}$
%bandwidth and transmit at $R_s^{in}$ and $R_f^{in}$ rates respectively. 
%In BE, the sender node transmits at $R
%and the forwarder node transmit with $W_s^{be}$ and $W_f^{be}$ bandwidth respectively. The sender nodes 
%The AP and the forwarder receive $R_d$ and $R_r$ data respectively from the sender. 
%The forwarder relays $R_c$ amount of sender's data along with transmitting its own data $R_f^{be}$ to the AP. The optimal resource allocation problem in this pair takes the following form:
%
\emph{\newline \underline{Problem IV}} 
\begin{subequations}
\begin{equation}
\mathbf{\max.} \, \,  U^{\alpha} (R_f^{be})  + U^{\alpha} (R_s^{be}) 
- U^{\alpha} (R_f^{in})  - U^{\alpha} (R_s^{in}) 
 \label{eq:NewObjective}
\end{equation}
\begin{eqnarray}
\mathbf{s.t.} \, \,  R_{sf} \leq W_s^{be} \log_2 \bigl(1 + \frac{P * \rho_{sf}}{W_s^{be}} \bigr) & & \nonumber \\  
R_{s0} \leq W_s^{be} \log_2 \bigl(1 + \frac{P * \rho_{s0}}{W_s^{be}} \bigr) & & \nonumber \\
R_c + R_f^{be} \leq W_f^{be} \log_2 \bigl(1 + \frac{P * \rho_{f0}}{W_f^{be}} \bigr) & & \label{eq:NewRateConstraint1}  
\end{eqnarray}
\begin{equation}
R_s^{be} \leq \min (R_{sf}, R_{s0} + R_c) \, , \, R_f^{be} \geq R_f^{in} \, , \, R_s^{be} \geq R_s^{in}   \label{eq:NewRateConstraint2}
\end{equation}
\begin{equation}
W_f^{be} + W_s^{be} \, \leq \, W_f^{in} + W_s^{in} \,  \, , \, \, W_f^{be}, \, W_s^{be} \, \geq 0 \, \, \forall \, (s,f) \, \in \, \mathcal{SF}  \label{eq:NewBandwidthConstraint}
\end{equation}
\begin{equation}
Variables \, \, \,  R_s^{be}, R_f^{be}, W_s^{be}, W_f^{be}, R_c \nonumber
\end{equation}
\end{subequations}
Equation~\eqref{eq:NewRateConstraint1} and~\eqref{eq:NewRateConstraint2}
show the convex hull of the allotted bandwidths and the achievable rates.
They have also already been described in the system model.
%Equation \eqref{eq:NewRateConstraint1} describes 
%that the data rates flowing in a path is upper bounded by the path capacities.
%Equation \eqref{eq:NewRateConstraint2} assumes
%that the sender-forwarder channel is better than the
%sender-AP channel, i.e., $\rho_{sf} \geq \rho_{s0}$, and hence, $R_r \geq R_d$.
%Therefor, the sender's rate in~\eqref{eq:NewRateConstraint2} is obtained by 
%using the max-flow-min-cut theorem~\cite{Cover}. 
%\eqref{eq:NewRateConstraint2} also denotes that the sender and forwarders' rates through BE cannot drop
%below their initial rates. The pairwise bandwidth constraint is represented in \eqref{eq:NewBandwidthConstraint}.

\underline{Lemmma 1:} Problem IV is a concave maximization problem.

\emph{Proof:} $U^{\alpha} (R_f^{in})$ and $U^{\alpha} (R_s^{in})$ are 
the utilites of the initial data rates and 
constants, in terms of the optimization variables. 
The concavity of $\alpha$-fair utility functions and 
the capacity expressions can be easily shown~\cite{Boyd}. The minimum
of linear (concave) functions is concave. 
Thus, the objective function is concave and the constraints are convex or linear in terms
of the optimization variables.
%Thus, the objective function is concave in terms of the bandwidths 
%and the constraints in \eqref{eq:NewRateConstraint1} \& \eqref{eq:NewRateConstraint2} are convex in terms %of bandwidths and rates.
%Besides, the constraints in \eqref{eq:NewBandwidthConstraint} is linear in terms of bandwidths. 
This proves the concavity of Problem IV.  $\blacksquare$.

Problem IV can be solved using standard convex optimization algorithms,
e.g., interior point methods~\cite{Boyd}.

%\underline{Lemma 2:} $(R_s^{be}, R_f^{be}, W_s^{be}, W_f^{be}, R_r, R_d, R_c) = (R_s^{in}, R_f^{in}, W_s^{in}, W_f^{in}, R_r^{in}, R_s^{in}, 0)$
%is a feasible solution of Problem IV where $R_r^{in} = W_s^{in} \log_2 \bigl(1 + \frac{P * \rho_{sf}}{W_s^{in}} \bigr)$.
%
%\emph{Proof:} Let $W_f^{be} = W_f^{in}$, $W_s^{be} = W_s^{in}$, $R_c = 0$. Then, 
%%
%\begin{eqnarray}
%& & R_r = W_s^{in} \log_2 \bigl(1 + \frac{P * \rho_{sf}}{W_s^{in}} \bigr) = R_r^{in} \nonumber \\ 
%& & R_d = W_s^{in} \log_2 \bigl(1 + \frac{P * \rho_{s0}}{W_s^{in}} \bigr) = R_s^{in}
%\end{eqnarray}
%%
%\begin{eqnarray}
%& & R_f^{be} = W_f^{in} \log_2 \bigl(1 + \frac{P * \rho_{f0}}{W_f^{in}} \bigr) = R_f^{in} \nonumber \\
%& & R_s^{be} = \min (R_r, R_d + R_c) = \min (R_r, R_d) = R_d = R_s^{in}
%\end{eqnarray}
%%
%This is a feasible solution of problem III. $\blacksquare$

\underline{Lemma 2:} $(R_s^{be}, R_f^{be}, W_s^{be}, W_f^{be}, R_c) = (R_s^{in}, R_f^{in}, W_s^{in}, W_f^{in}, 0)$
is a feasible set of variables of Problem IV.

\emph{Proof:} Let $W_f^{be} = W_f^{in}$, $W_s^{be} = W_s^{in}$, $R_c = 0$. Then, 
\begin{eqnarray}
& & R_{sf} = W_s^{in} \log_2 \bigl(1 + \frac{P * \rho_{sf}}{W_s^{in}} \bigr)  \nonumber \\ 
& & R_{s0} = W_s^{in} \log_2 \bigl(1 + \frac{P * \rho_{s0}}{W_s^{in}} \bigr) = R_s^{in}
\end{eqnarray}
\begin{eqnarray}
& & R_f^{be} = W_f^{in} \log_2 \bigl(1 + \frac{P * \rho_{f0}}{W_f^{in}} \bigr) = R_f^{in} \nonumber \\
& & R_s^{be} = \min (R_{sf}, R_{s0} + R_c) = \min (R_{sf}, R_{s0}) = R_s^{in}  \nonumber
\end{eqnarray}
This is a feasible solution of problem IV. $\blacksquare$

\begin{figure*}[t]
\centering
\includegraphics[scale=0.55]{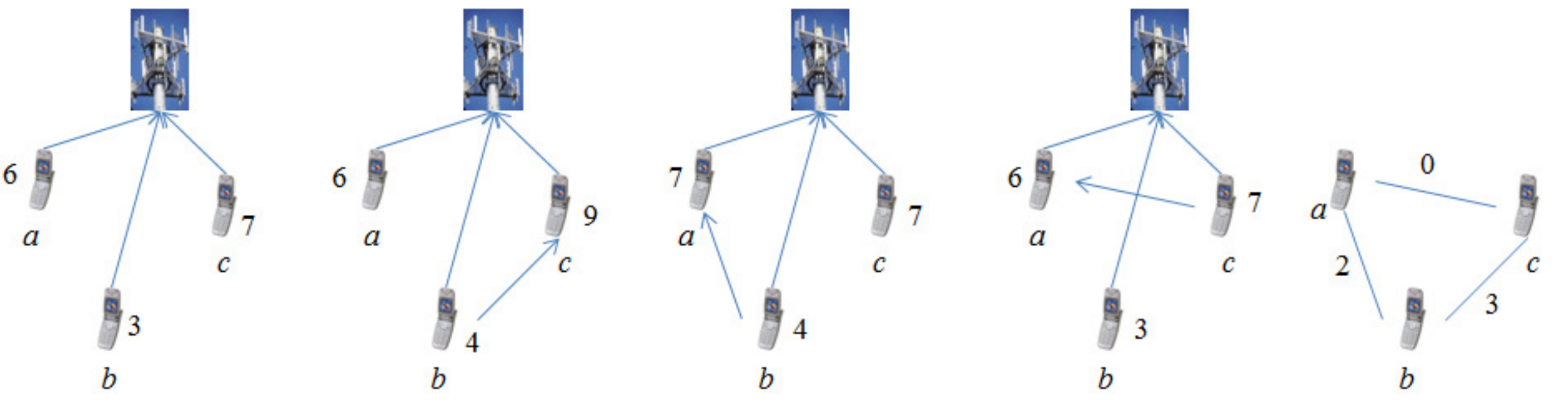}
\caption{MWM in BE Enabled Relay Network}
\label{fig:MWMIllustration}
\end{figure*}

Thus, if Node $s$ and $f$ use their initial bandwidths and if the forwarder $f$ does not relay any data of the sender $s$, both 
sender and forwarder will continue to transmit at their initial rates.
% In other words, the initial bandwidth allocation and
%direct path transmission based rates are also feasible solutions for BE enabled incentivized forwarding. 
The corresponding feasible solution for these variables can be found as 
follows,
\begin{eqnarray}
& & U^{\alpha} (R_f^{be})  + U^{\alpha} (R_s^{be}) - U^{\alpha} (R_f^{in})  - U^{\alpha} (R_s^{in})
\nonumber \\ 
& & = U^{\alpha} (R_f^{in})  + U^{\alpha} (R_s^{in}) - U^{\alpha} (R_f^{in})  - U^{\alpha} (R_s^{in}) = 0 \nonumber
\end{eqnarray}
Since, problem IV is a concave maximization probelm, the optimal solution will not drop below $0$~\cite{Boyd}.
%Hence, $ U^{\alpha} (R_f^{be})  + U^{\alpha} (R_s^{be}) \, \geq \, U^{\alpha} (R_f^{in})  + U^{\alpha} (R_s^{in})$.
Hence, the proposed BE enabled relaying scheme will perform at least as good as
the initial allocation and \emph{seek to maximize the global utility
while preserving the initial rates of the individual nodes}.
%Therefore, the proposed half duplex DF cooperative relaying will maximize the summation of the utilites %of the cooperative
%pair while ensuring that the individual nodes' data rates don't drop below their initial ones.
%Simulation results in Sec.~\ref{sec:Simulation} will validate these observations.

%Bandwidth allocation through a similar half duplex DF mechanism previously
%appreared in~\cite{Foschini}. However, the authors of~\cite{Foschini} considered a commercial relay network
%where the relay did not have its own data to transmit and focused entirely on sender's rate maximization.

\underline{Lemma 3:} Problem III is a concave maximization problem.

\emph{Proof:} Problem III is the combination of $K$ disjoint concave maximization
problems and hence it's concave.  $\blacksquare$

\subsection{Optimal sender-forwarder set selection}

Let $g_{sf}^*$ denote the solution of problem IV, i.e.,
the optimal gain obtained through cooperation of $(s, \, f)$.
Let, $U_{\mathcal{D},\mathcal{SF}}^*$ represent the optimal solution
of problem III, i.e., the optimal cooperation gain for the selected sender-forwarder pair set
$\mathcal{SF}$. Then, $U_{\mathcal{D},\mathcal{SF}}^* =  \sum_{(s, \, f) \in \mathcal{SF}} g_{sf}^*$.
Therefore, the sender-forwarder pairing set selection part of problem II 
is equivalent to finding the set of pairs that maximize the gain
in utility through cooperation, over noncooperation. It can be written in
the following form:

\emph{\newline \underline{Problem V}}
 \begin{subequations}
\begin{equation}
\displaystyle
\mathbf{\arg \max_{\mathcal{SF}}} \sum_{(s, \, f) \in \mathcal{SF}} g_{sf}^* 
\label{eq:BEMasterOpt}
\end{equation}
\begin{equation}
\mathcal{SF} \in \mathcal{V} \times \mathcal{V} \, \, , \, \,
\mathcal{SF}_i \cap \mathcal{SF}_j = \emptyset  \, \, \forall i \neq j \, , \, \label{eq:SubRelaySelection1}
%\mathcal{SF} \cap \mathcal{D} = \emptyset \, , \, \mathcal{SF} \cup \mathcal{D} = \mathcal{V}  \label{eq:SubRelaySelection1}
\end{equation}
\end{subequations}
%

%Now, the solution of Problem IV is the difference between the cooperation and
% non-cooperation utility of the $(s, f)$ pair.
Now, consider a graph $ \mathcal{G} = (\mathcal{V},\mathcal{E})$ where the vertices $\mathcal{V}$
represent the set of $N$ nodes under consideration and $\mathcal{E}$ denote the edges between these nodes.
Define the edge weight of any $(i,j)$ pair by $U^{\alpha} (R_i^{be})  +  U^{\alpha} (R_j^{be})
- U^{\alpha} (R_i^{in})  -  U^{\alpha} (R_j^{in})$, i.e., the difference, in terms of utility, 
between the cooperation and non-cooperation scenario. 
The optimal sender-forwarder pairing set selection of problem V
is equivalent to finding the set of pairs that 
maximize the difference between cooperation and noncooperation utility. 
Hence, the optimal sender-forwarder pair selection problem 
can be reduced to the problem of finding the set of pairs that maximizes the link weights
mentioned above. Thus, the optimal relay selection converges to
the classical nonbipartite MWM problem. The 
nonbipartite MWM algorithm has been summarized in Appendix~\ref{sec:Matching}.
A detailed description can be found in~\cite{Edmonds}. 
%We skip the details here due to space constraints.
%Appendix A briefly describes the optimal MWM in a non-bipartite graph problem 
%formulation.

Fig.~\ref{fig:MWMIllustration} illustrates the application of MWM 
in the sender-forwarder pairing selection.
The left figure of Fig.~\ref{fig:MWMIllustration} denote the initial scenario where node $a$, $b$
and $c$ transmit through the direct path and transmit $6$, $3$ and $7$ bits respectively.
The three figures in the middle show the obtained rates for different sender-forwarder pair selection.
The second figure from the left shows that $b$ and $c$ transmit $4$ and $9$ bits respectively through
BE enabled DF cooperation. Thus, the utility gain of cooperation, over noncooperation, is $3$ bits.
The middle figure and the 2nd figure from the right represent the cooperation scenarios of $(a,b)$
and $(a,c)$ respectively. The rightmost figure represents the edge weights of each cooperative pair
in terms of the utility gain of cooperation, over noncooperation. The MWM algorithm
will select node $(b,c)$ as the cooperative pair and node $a$ will transmit without cooperation.

Centralized nonbipartite MWM can be solved optimally in $O(N^3)$ time~\cite{Edmonds}.
%Hoepman~\cite{Hoepman} developed a distributed local greedy MWM algorithm to find the matching pairs. 
Our proposed distributed incentivized forwarding is based on the distributed local
greedy MWM~\cite{Hoepman} $(O(N^2))$
and is described below. The distributed MWM algorithm~\cite{Hoepman}
is summarized in Appendix~\ref{sec:DistMWM}.
%We skip the details of the protocol here due to space constraints.
%The final version
%of the paper, if accepted, will contain the protocol.

\subsection{Distributed BE incentivized forwarding protocol}

\begin{itemize}

\item Focus on an arbitrary node, node $v$. Node $v$ sends training symbols to the AP and obtains its
own direct channel, $\rho_{v0}$, through feedback. Node $v$ is initially allocated $W_v^{in}$ bandwidth
and transmits at $R_v^{in}$ rate.

\item Let node $u$ be a neighbouring node of $v$. Due to the nature of wireless channels, node $u$ receives
node $v$'s channel estimation training symbols and finds the inter-node channel gain, $\rho_{uv}$.

\item $v$ sends an omnidirectional signal containing $\rho_{v0}$, $W_v^{in}$ and $R_v^{in}$ to the neighbouring nodes.

%\item Each node follows the three steps mentioned above.
% In this way, each node contains 
%the necessary parameters to solve problem IV.

\item Node $u$ may relay node $v$'s data if $min(\rho_{uv},\rho_{u0}) \geq \rho_{v0}$.
Thus, $v$ knows its potential forwarders or senders.
%Based on this, node $v$ has three list of neighbouring nodes. First, potential senders for $v$.
%Second, potential forwarders for $v$. Third, the other nodes that can neither act as
%forwarders or senders for $v$.
 
\item $v$ solves problem IV for the suitable neighbours. 
Thus, each node knows its adjacent link weights.

\item $v$ solves the distributed local greedy MWM algorithm of~\cite{Hoepman}.
 %$v$ can find its ``candidate pair node" in the following two ways: \\
 %1. Solve the distributed local greedy MWM algorithm of~\cite{Hoepman}. \newline
% 2. If the channel environment is very dynamic, then 
% nodes cannot afford the waiting time involved with the algorithm of~\cite{Hoepman}.
% In this case, node $u$ and $v$ form a pair only if they are the best candidate nodes of each other.
% i.e., if the link weight of $(u,v)$ is locally maximum to both $u$ and $v$.
 
%\item Once relays are selected, channel estimation between sender-forwarder pair occurs. 
%Inter-node link gain history is updated.
\item The `matched pairs' allocate resources among themselves. The `unmatched' nodes transmit without
cooperation.
% set of nodes
%that do not find a cooperative pair transmit in the direct path with their primary resources.

\end{itemize} 

\subsection{Outage probability reduction in BE}

We define outage probability as the ratio of the number of
nodes that do not get minimum data rate to the total number
of nodes. We assume that each node in the network starts with
an initial amount of resource. Depending on the resource and
link gains, nodes fall in the following two groups:

\begin{itemize}
\item Outage group: Node that cannot meet the minimum
required rate with initially allocated resources.
\item Non-Outage group: Node that can meet the minimum
rate with initially allocated resources.
\end{itemize}

The outage probability reduction problem can be defined as
providing minimum data rate to the most number of users in
the outage group, while maintaining the minimum data rate
of the nodes, in the non-outage group. MWM based matching
and pairwise resource allocation based incentivized two-hop
forwarding can help in this case. We propose the following
scheme in this regard:

\begin{itemize}

\item Each node in the outage group solves the pairwise
sumrate maximization, with minimum rate constraints,
for each of its neighbouring node. 
%We use sumrate maximization
%with minimum rate constraints since it provides
%the most benefit to the forwarder while maintaining
%the sender’s minimum data rate~\cite{Chiang:a}. 
Nodes can solve sumrate maximization by plugging $\alpha = 0$ in the $\alpha$-fair
utility function.

\item If the node in outage can maintain minimum
data rate by pairing with the forwarder, i.e., the non-outage
node, we assume that an edge exists between these
nodes.

\item The relay selection problem in outage probability reduction
becomes maximizing the number of edges in the
network. This reduces to the maximum matching (MM) algorithm in a
bipartite graph~\cite{Nemhauser}.

\end{itemize}

Our focus here is to maximize the number of users that
receive minimum data rate, not to maximize any utility function.
That's why, we use MM, instead of MWM, in this part
of the work. Besides, cooperation between two nodes in the
non-outage group do not change the outage probability of the
network. Therefore, we consider a bipartite graph by dividing
the graph into outage and non-outage group.

\section{Numerical Simulations}   \label{sec:Simulation}
\begin{figure}[t]
%\begin{minipage}[b]{0.5\linewidth}
\centering
\includegraphics[scale=0.45]{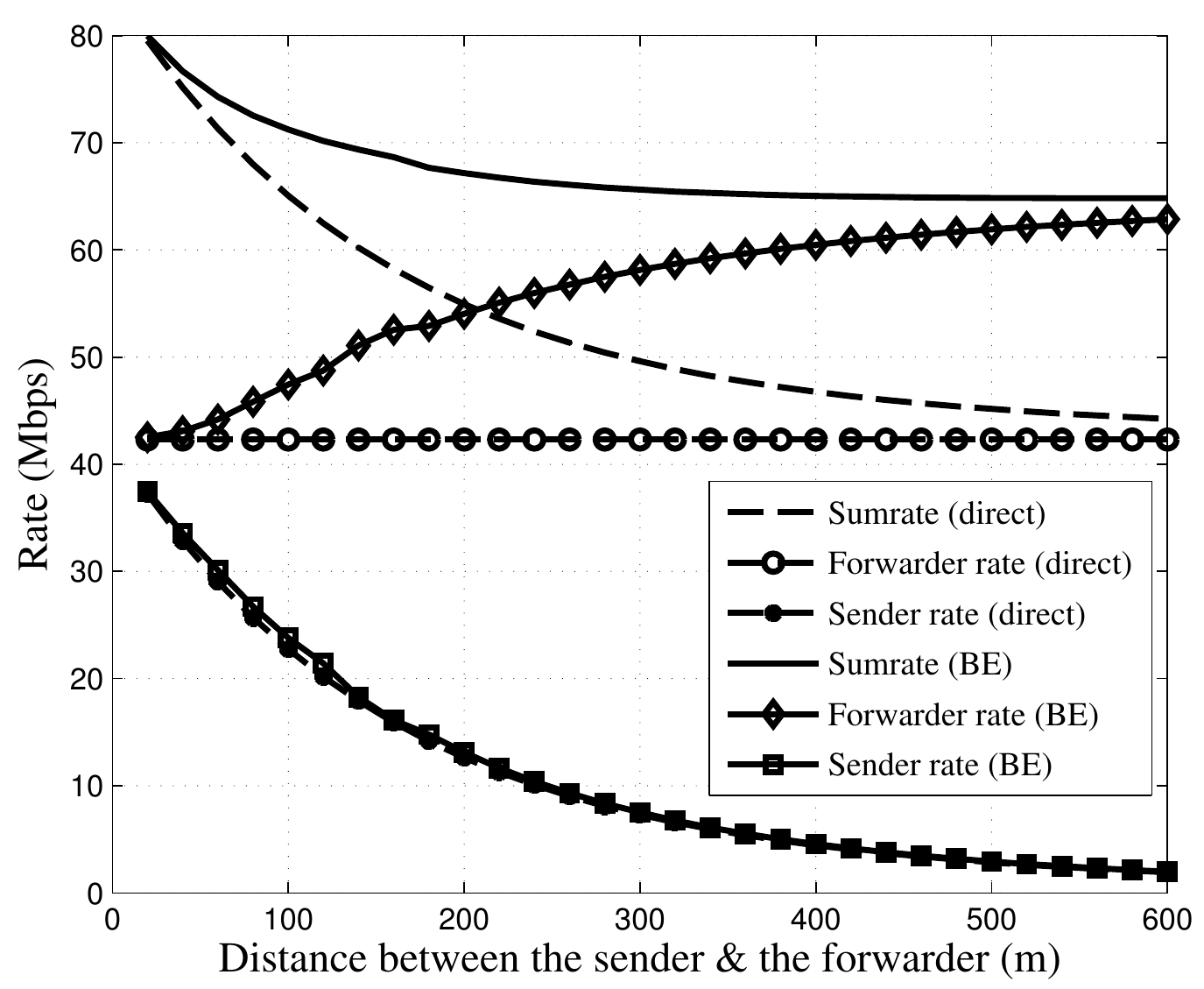}
\caption{Sumrate Maximization in a 3 Node Network, $10$ MHz per Node, $P$ = $100$ mW, Near node-AP distance = 150m}
\label{fig:MaxSumRateTwoNode}
\end{figure}
%\end{minipage}
%\hspace{0.75cm}
%\begin{minipage}[b]{0.5\linewidth}
\begin{figure}
\centering
\includegraphics[scale=0.45]{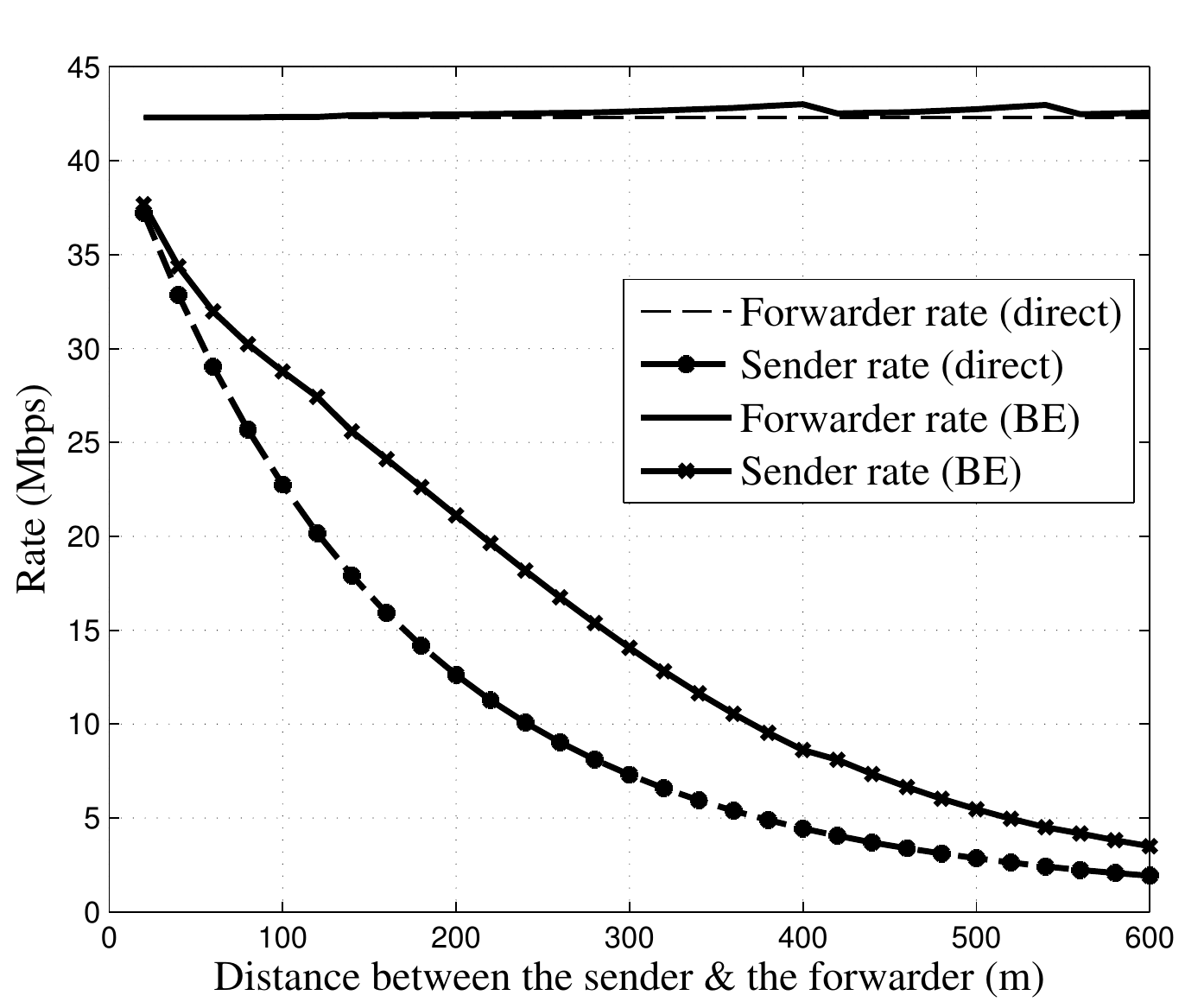}
\caption{Minimum Rate Maximization in a 3 Node Network, $10$ MHz per Node, $P$ = $100$ mW, Near node-AP distance = 150m}
\label{fig:MaxMinRateTwoNode}
%\end{minipage}
\end{figure}
We assume equal initial bandwidth allocation in all of our simulations, i.e., 
nodes start with equal bandwidth. However, our work is readily applicable to the scenario where nodes start with optimal bandwidth allocation (based on direct path transmission) and then
use bandwidth as incentives for two hop relaying.

%We first show numerical results to illustrate the performance of BE enabled resource allocation
%strategy in a $3$ node work. 
%The plots of Fig.~\ref{fig:MaxSumRateTwoNode} and Fig.~\ref{fig:MaxMinRateTwoNode}
%are based on a network where $2$ nodes transmit uplink data to the AP.
%The near  
Fig.~\ref{fig:MaxSumRateTwoNode} and Fig.~\ref{fig:MaxMinRateTwoNode}
compare the performance of BE relaying with that of direct path transmission
in the sumrate maximization and minimum rate maximization of a $3$ node
network (sender, forwarder and BS). Both sender and forwarder initially receive 10 MHz 
bandwidth and transmit uplink data to the BS. 
The forwarder node is placed in the straight line that connects the BS and the sender node.
The distance between the forwarder node and the BS is kept fixed at $150$m, whereas, the distance between 
the BS \& the sender node is varied. 
%Two nodes received $20$ MHz bandwidth in total.
In these two simulations, we assumed the link gains to take the form, $\rho_{ij} = k d_{ij}^{-3}$
where $d_{ij}$ is the distance between the $i^\mathrm{th}$ and $j^\mathrm{th}$ node.
%We also assume, $\rho_{ij} = \rho_{ji}$. 
$k$ is the proportionality constant that also captures the
noise spectral density and is set to $k = 6 \times 10^6 MHz * m^3/mW$~\cite{Zhang}.

\begin{figure}[t]
%\begin{minipage}[b]{0.5\linewidth}
\centering
\includegraphics[scale=0.45]{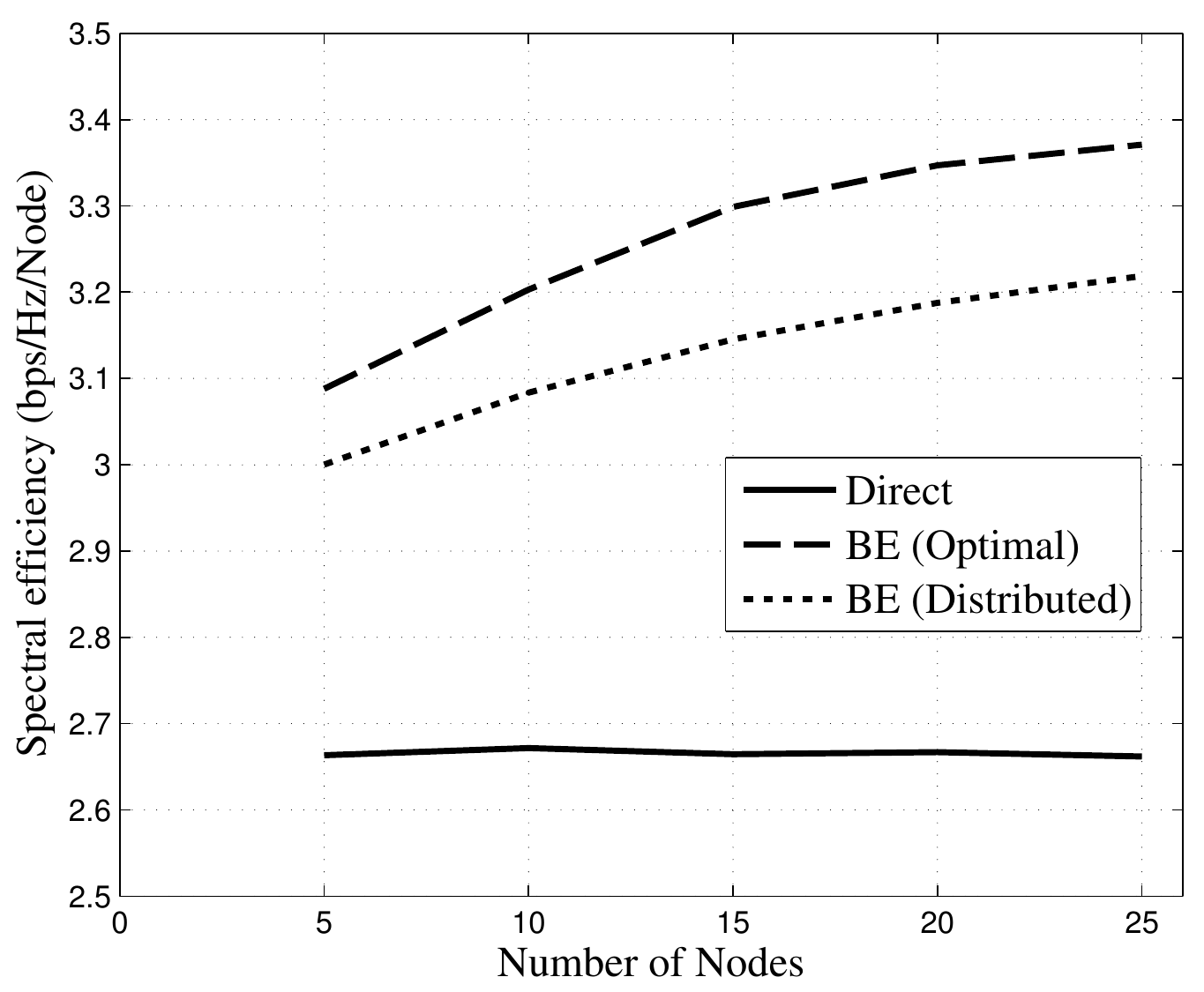}
\caption{Spectrum Efficiency in an $N$ Node Network, $1$ MHz per Node, $P$ = $20$ dBm }
\label{fig:SpectrumEfficiency}
%\end{minipage}
\end{figure}
\hspace{1cm}
%\begin{minipage}[b]{0.5\linewidth}
\begin{figure}[t]
\centering
\includegraphics[scale=0.45]{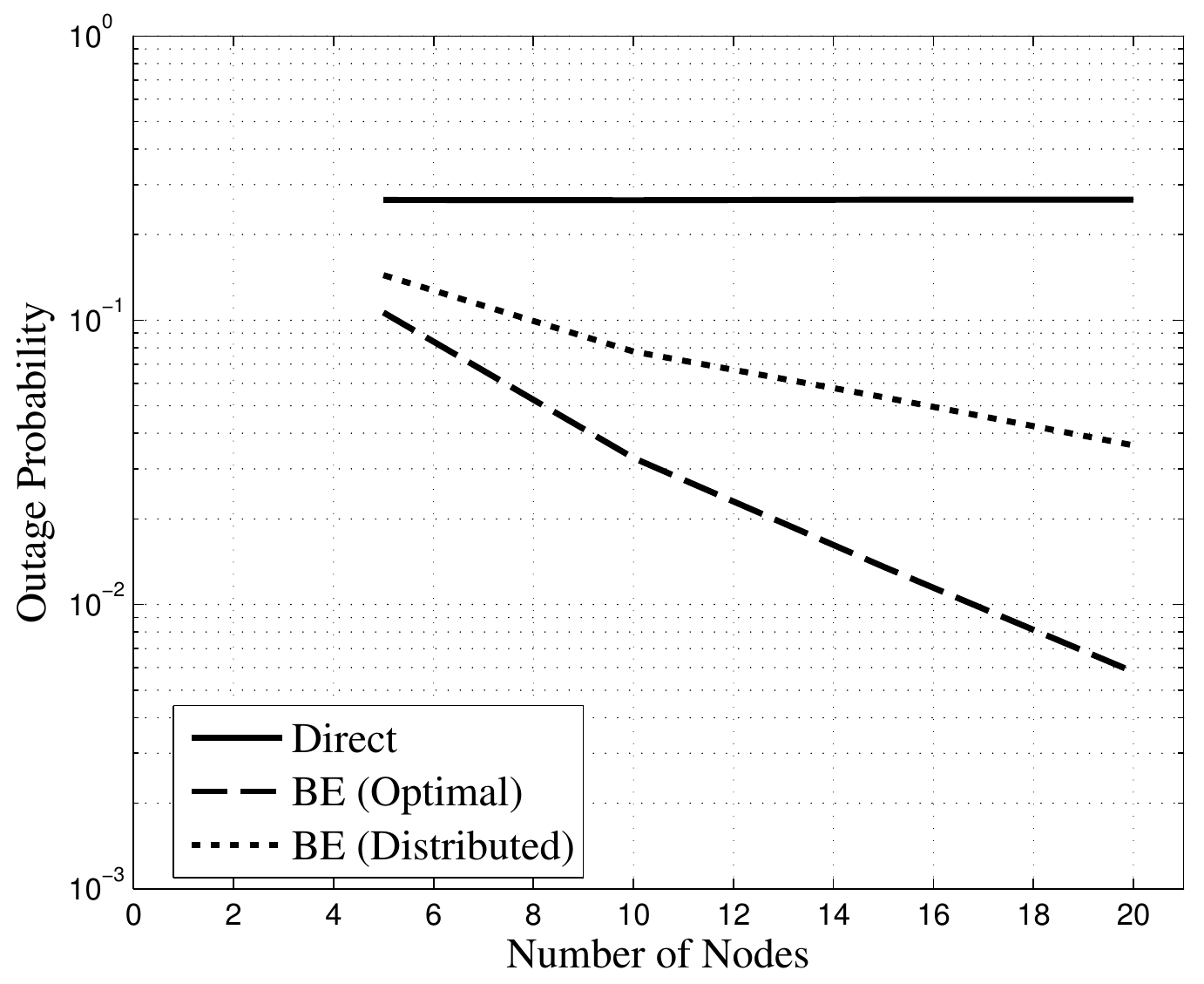}
\caption{Outage Probability in an $N$ Node Network, $1$ MHz per Node, $P$ = $20$ dBm }
\label{fig:OutageProbability}
%\end{minipage}
\end{figure}

The sumrate maximization objective based plot of 
Fig.~\ref{fig:MaxSumRateTwoNode} shows that BE relaying improves the rate of the
forwarder (near user) while ensuring that the sender's (far user) rate does not drop below its
initial value. This increase in the forwarders' rate gets reflected in the sumrate gain of the network.
On the other hand, the minimum rate maximization objective based plot of Fig.~\ref{fig:MaxMinRateTwoNode}
shows that BE relaying improves the senders' (far user) rate while ensuring that the forwarders' (near user)
rate does not drop. This diverse contribution of BE relaying comes from the 
problem objectives of the respective simulations; sumrate maximization ($\alpha = 0$)
is the most efficient allocation whereas minimum rate maximization ($\alpha = \infty$) is the most fair one.
Therefore, the use of any $\alpha \in (0,\infty)$ would have increased both users' rates
in this simulation scenario.

Fig.~\ref{fig:SpectrumEfficiency} and Fig.~\ref{fig:OutageProbability} show the performance
of BE relaying in an $N$ node network. In these simulations, we assumed that links
are under independent Rayleigh fading and the link gain in each slot is an independent
realization of Rayleigh random variable. This means that the link gain, $\rho_{ij}$,
is exponentially distributed, 
$p(\rho_{ij}) = \frac{1}{\bar{\rho}_{ij}} exp \bigl( - \frac{\rho_{ij}}{\bar{\rho}_{ij}} \bigr)$
where $\bar{\rho}_{ij} = k d_{ij}^{-3}$ and $k = 6 \times 10^6 MHz * m^3/mW$.
We consider a circular cell of $800m$ radius. The AP is located at the center, whereas,
the nodes are placed randomly in the cell. 
We considered transmission scheme much like the one used in mobile Wimax.
Each node is preassigned 20 dBm transmit power and 1 MHz bandwidth.
We used the matching code of~\cite{Joris} to implement the MWM
algorithm in Matlab.

We showed the performance of both centralized and distributed algorithms in Fig.~\ref{fig:SpectrumEfficiency}
and Fig.~\ref{fig:OutageProbability}. In the simulation of the distributed algorithm,
we assumed that each node can only talk to its neighbours that are located within $500$m
from the sender node. Fig.~\ref{fig:SpectrumEfficiency} shows that the 
centralized and decentralized algorithm improves the spectral efficiency by 25\% and 20\%
respectively. The performance of the distributed algorithm will improve 
if we allow each node to talk to neighbours with greater distances.

Fig.~\ref{fig:OutageProbability} shows that BE enabled relaying provides
cooperative diversity and 
significantly reduces the outage probability (90-98\%). Thus, BE can be used
to extend the coverage in an autonomous network.

\section{Discussion}   \label{sec:Conclusion}

In this paper, we considered joint optimal relay selection and resource allocation
in the $\alpha$-fair NUM and outage probability reduction 
of a BE network. Our proposed resource allocation formulation maximizes the global utility
of the cooperative pair while preserving the initial utilities
of each individual node. We showed that the relay selection part
of the  $\alpha$-fair NUM problem reduces to the nonbipartite matching
algorithm. Numerical simulations suggest that the proposed BE enabled relaying
provides 20-25\% spectrum efficiency gain and 90-98\% outage probability reduction in a
$20$ node network. 
%The final version of the algorithm will show the transmission 
%power savings of the proposed algorithm and its application in a TDMA network. 

Our work can be really advantageous in the following scenario:
the nodes start with a direct path based centrally allocated optimal resources.
Thereafter, nodes employ the proposed algorithm to improve the system performance
through distributed relay selection and pairwise resource allocation.
This scheme saves a huge amount of signalling overhead, an inherent drawback
of centralized optimal two hop forwarding, at the cost of lower performance.

The proposed algorithm considers one forwarder for one sender and vice versa.
The generalization of this algorithm to the multiple sender-forwarder
scenario is an area of future research. We also plan to focus on discrete subcarrier
exchange in our future works.

\section{Acknowledgements}

This work is supported by the Office of Naval Research under grant N00014-11-1-0132.
We thank Vladimir Oudalov and Dr. Mung Chiang for their insights on
maximum weighted matching algorithm. We are also grateful to Joris V. Rantwijk
for the MWM implementation code~\cite{Joris}.

\begin{appendices}

\section{Codebook Design in Proposed DF Relaying}   \label{sec:Codebook}

Let's consider two codebooks $\mathcal{W}$ and $\mathcal{C}$ that
consist of $2^{R_{sf}}$ and $2^{R_c}$ codewords respectively.
Assume, $\mathcal{W} = \{w_1, w_2, \cdots, w_{2^{R_{sf}}}\}$ 
and $\mathcal{C} = \{c_1, c_2, \cdots, c_{2^{R_c}}\}$. 
Here $R_{sf} \geq R_c$. Consider a partition
$\mathcal{S} = \{S_1, S_2, \cdots, S_{2^{R_c}}\}$ of $\mathcal{W}$, i.e.,
$\mathcal{W}$ has been partitioned into $2^{R_c}$ cells.
Each cell $S_i$ contains $2^{R_{sf}-R_c}$ codewords of $\mathcal{W}$.
%Here, $S_i \cap S_j = \emptyset \,, \, i \neq j \,, \, \cup S_i = \mathcal{W}$.
Assume a one-to-one correspondence between $\mathcal{C}$ and $\mathcal{S}$, i.e.,
each codeword of $\mathcal{C}$ represents one particular cell of $\mathcal{S}$.

The BS (node $0$), sender $s$ and forwarder $f$ 
get the codebooks off-line. At the beginning of
transmission, sender $s$ sends a codeword $w_i$ from $\mathcal{W}$
using $R_{sf}$ bits. The forwarder node decodes the codeword correctly. However, since
$R_{sf} \geq R_{s0}$, the BS cannot decode it correctly. The BS 
has a list of possible codewords of size $2^{R_{sf} - R_{s0}}$. 
Now, the forwarder $f$ finds the cell
$S_i$ where $w_i$ lies and sends $c_i$ using $R_c$ bits. The BS receives $c_i$
and intersects $S_i$ with the list of possible codewords. If $R_c \geq R_{sf} - R_{s0}$
and $R_c \leq R_{f0}$, this half duplex DF cooperation completely
removes the BS's uncertainty about $w_i$~\cite{Cover,Cover2}.

Thus, the achievable rates of node $s$ and $f$ are governed by this information 
theoretic generalization of the max-flow-min-cut theorem:
\begin{eqnarray}
R_s^{be} & \leq & \min (R_{sf}, R_{s0} + R_c)  \nonumber    \\
R_c + R_f^{be} & \leq & R_{f0}     \nonumber
\end{eqnarray}

\section{Matching and nonbipartite MWM algorithm}  \label{sec:Matching}

Consider an undirected graph $\mathcal{G} = (\mathcal{V}, \mathcal{E})$ where
$\mathcal{V}$ denotes the set of vertices and $\mathcal{E}$ denotes the set of edges. A matching
$\mathcal{M}$ is a subset of $\mathcal{E}$ such that 
$ e_1 \cap e_2 = \emptyset$  for $e_1, e_2 \in \mathcal{M}$ if $e_1 \neq e_2$~\cite{Nemhauser}.

Let $x_e$ denote whether an edge $e \in E$ will be selected in the matching, i.e., $x_e$
can be $0$ or $1$. Let $r_e$ represent the edge weights.
The maximum weighted matching in a non-bipartite graph takes the 
following form~\cite{Nemhauser}:
\begin{subequations}
\begin{equation}
%\displaymath
\mathbf{max} \sum_{e \in E} r_e x_e  \label{eq:OptBipartite}
\end{equation}
\begin{equation}
%\displaymath
\mathbf{s.t.} \sum_{e \in \delta(v)} x_e \leq 1 \, \forall \, v \, \in \, \mathcal{V}  \label{eq:BiPartite1}
\end{equation}
\begin{equation}
%\displaymath
\sum_{e \in E(\mathcal{U})} x_e \leq \lfloor \frac{|\mathcal{U}|}{2} \rfloor \, \forall \, \, odd \, sets \, \, \mathcal{U} \subset \mathcal{V} 
\label{eq:BiPartite2}
\end{equation}
\begin{equation}
x_e \in \mathcal{B}^n  \label{eq:BiPartite3}
\end{equation}
\end{subequations}
In \eqref{eq:BiPartite1}, $\delta (v)$ denotes the edges connected with node $v$.
In \eqref{eq:BiPartite2}, $E(\mathcal{U})$ represents the edges contained in the set $\mathbf{U}$.
Equation \eqref{eq:BiPartite3} shows that $x_e$ takes Boolean values.
However, Edmonds~\cite{Edmonds} showed that the we can replace $x_e \in \mathcal{B}^n$
by $x_e \in \mathcal{R}^n_+$ and still obtain integral optimal solutions.
Thus the combinatorial optimization can be converted to a linear program.

\section{Distributed Local Greedy MWM}  \label{sec:DistMWM}

\begin{itemize}

\item Each node $i$ knows its adjacent link weights. Node $i$
picks the ``candidate'' node $j$, based on the heaviest link
weight and sends an ``add'' request.
\item Wait for the response from node $j$.
\item If node i receives an ``add'' request from node $j$,
$i$ and $j$ pick each other as the cooperative pair. $i$ sends
``drop'' request to its other neighbouring nodes.
\item If node i receives a ``drop'' request from node $j$,
node $i$ removes the $(i, j)$ link from its adjacent edge set.
Node $i$ goes to the state of step $1$.

\end{itemize}

The distributed local greedy MWM  provides at least
50\% performance of centralized optimal matching. Distributed
local greedy MWM requires $O(N^2)$ amount of message
passing.

\end{appendices}

\bibliographystyle{IEEEbib}
\bibliography{BibWiOpt}

\begin{thebibliography}{10}

\bibitem{Tse}
J.~N. Laneman, D.~N.~C. Tse, and G.~Wornell,
\newblock ``Cooperative diversity in wireless networks : efficient protocols
  and outage behavior,''
\newblock {\em IEEE Trans. Info. Theory}, vol. 50(12), pp. 3062--3080, Dec.
  2004.

\bibitem{ElGamal}
L.~Lai, K.~Liu, and H.~El Gamal,
\newblock ``The three node wireless network: Achievable rates and cooperation
  strategies,''
\newblock {\em IEEE Trans. Info. Theory}, vol. 52(3), pp. 805--828, Mar. 2006.

\bibitem{Ileri}
O.~Ileri, S.-C. Mau, and N.~Mandayam,
\newblock ``Pricing for enabling forwarding in self-configuring ad hoc
  networks,''
\newblock {\em IEEE JSAC}, vol. 23, pp. 151--162, Jan. 2005.

\bibitem{Buchegger}
S.~Buchegger and J.-Y~Le Boudec,
\newblock ``Self-policing mobile ad hoc networks by reputation systems,''
\newblock {\em IEEE Communication Magazine}, vol. 43, pp. 101--107, July 2007.

\bibitem{Buttyan}
M.~Felegyhazi, J.~P. Hubaux, and L.~Buttyan,
\newblock ``Nash equilibria of packet forwarding strategies in wireless ad hoc
  networks,''
\newblock {\em IEEE Trans. on Mobile Computing}, vol. 5, pp. 463--475, May
  2006.

\bibitem{Zhang}
D.~Zhang, R.~Shinkuma, and N.~B. Mandayam,
\newblock ``Bandwidth exchange: An energy conserving incentive mechanism for
  cooperation,''
\newblock {\em IEEE Trans. Wireless Comm}, vol. 9(6), pp. 2055--2065, June
  2010.

\bibitem{Edmonds}
J.~Edmonds,
\newblock ``Paths, trees and flowers,''
\newblock {\em Canadian Journal of Mathematics}, vol. 17, pp. 449--467, 1965.

\bibitem{Hoepman}
J.~Hoepman,
\newblock ``Simple distributed weighted matchings,'' eprint, October 2004,
\newblock http://arxiv.org/abs/cs/0410047.

\bibitem{Zhang2}
J.~Zhang and Q.~Zhang,
\newblock ``Stackelberg game for utility-based cooperative cognitive radio
  networks,''
\newblock in {\em Proc. {ACM} {MOBIHOC}'2009}, May 2009, pp. 23--31.

\bibitem{Baochun:a}
H.~Xu and B.~Li,
\newblock ``Efficient resource allocation with flexible channel cooperation in
  ofdma cognitive radio networks,''
\newblock in {\em Proc. {IEEE} {INFOCOM}'2010}, Mar. 2010, pp. 1--9.

\bibitem{Nazmul}
M.~N. Islam, N.~B. Mandayam, and S.~Komplella,
\newblock ``Optimal resource allocation in a bandwidth exchange enabled relay
  network,''
\newblock in {\em Proc. {IEEE} {MILCOM}'2011}, Nov. 2011.

\bibitem{Foschini}
C.~T.~K. Ng and G.~J. Foschini,
\newblock ``Transmit signal and bandwidth optimization in multiple-antenna
  relay channels,''
\newblock {\em IEEE Trans. Wireless Comm.}, vol. 59, pp. 2987--2992, Nov. 2011.

\bibitem{Ephremides}
L.~Tassiulas and A.~Ephremides,
\newblock ``Stability properties of constrained queing systems and scheduling
  for maximum throughput in multihop radio networks,''
\newblock {\em IEEE Trans. Automatic Control}, vol. 37(12), pp. 1936--1949,
  Dec. 1992.

\bibitem{Neely}
M.~J. Neely, E.~Modiano, and C.~E. Rohrs,
\newblock ``Dynamic power allocation and routing for time-varying wireless
  networks,''
\newblock {\em {IEEE} Journal on Selected Areas in Communications}, vol. 23(1),
  pp. 89--104, Jan. 2005.

\bibitem{Saswati}
S.~Sarkar and L.~Tassiulas,
\newblock ``End-to-end bandwidth gurantees through fair local spectrum share in
  wireless ad-hoc networks,''
\newblock in {\em Proc. {IEEE} Conference on Decision and Control}, Dec. 2011.

\bibitem{Chiang:b}
Y.~Yi and M.~Chiang,
\newblock ``Stochastic network utility mazmimization and wireless scheduling,''
\newblock in {\em Next Generation Internet Architectures and Protocols},
  B.~Ramamurthy, G.~Rouskas, and K.~Sivalingam, Eds., pp. 1--35. Cambridge
  University Press, New York, 2011.

\bibitem{Shroff}
X.~Lin and N.~B. Shroff,
\newblock ``The impact of imperfect scheduling on cross-layer rate control in
  wireless networks,''
\newblock in {\em Proc. {IEEE} {INFOCOM} 2005}, Mar. 2005.

\bibitem{Mak}
V.~Mahinthan, J.~Mark L.~Cai, and X.~Shen,
\newblock ``Maximizing cooperative diversity energy gain for wireless
  networks,''
\newblock {\em IEEE Trans. Wireless Comm}, vol. 7(6), pp. 2540--2549, 2007.

\bibitem{Cover2}
T.~Cover and H.~El Gamal,
\newblock ``Capacity theorems for the relay channel,''
\newblock {\em IEEE Trans. Info. Theory}, vol. 25(5), pp. 572--584, Sept. 1979.

\bibitem{Cover}
T.~M. Cover and J.~A. Thomas,
\newblock {\em Elements of Information Theory},
\newblock John Wiley and Sons, Hoboken, NJ, 2005.

\bibitem{Mo}
J.~Mo and J.~Warland,
\newblock ``Fair end-to-end window based congestion control,''
\newblock {\em IEEE/ACM Transactions on Networking}, vol. 8(5), pp. 556--567,
  Oct. 2000.

\bibitem{Boyd}
S.~Boyd and L.~Vandenberghe,
\newblock {\em Convex Optimization},
\newblock Cambridge University, Cambridge, UK, 2004.

\bibitem{Nemhauser}
G.~Nemhauser and L.~Wolsey,
\newblock {\em Integer and Combinatorial Optimization},
\newblock John Wiley and Sons, Hoboken, NJ, 1988.

\bibitem{Joris}
Joris van Rantwijk,
\newblock ``Maximum weighted matching,''
  http://jorisvr.nl/maximummatching.html/, accessed February 2012.

\end{thebibliography}

\end{document}